\documentclass{rsauthor}
\usepackage[T1]{fontenc}
\usepackage[latin9]{inputenc}
\usepackage{babel}
\usepackage{textcomp}
\usepackage{mathrsfs}
\usepackage{amsmath}
\usepackage{amssymb}
\usepackage{graphicx}

\usepackage{amsfonts}\usepackage{subfigure}

\providecommand{\U}[1]{\protect\rule{.1in}{.1in}}

\jname{Phil. Trans. R. Soc. A}

\begin{document}

\title{On the quantumness of correlations in nuclear magnetic resonance}

\author{D. O. Soares-Pinto$^{1}$, R. Auccaise$^{2}$, J. Maziero$^{3}$, A. Gavini-Viana$^{4}$, R. M. Serra$^{3}$ and L. C. C\'{e}leri$^{3}$}
\address{$^{1}$Instituto de F\'{\i}sica de S\~{a}o Carlos, Universidade de S\~{a}o Paulo, Caixa Postal 369, 13560-970 S\~{a}o Carlos, S\~{a}o Paulo, Brazil \\
$^{2}$Empresa Brasileira de Pesquisa Agropecu\'{a}ria, Rua Jardim Bot\^{a}nico 1024, 22460-000 Rio de Janeiro, Rio de Janeiro, Brazil \\
$^{3}$Centro de Ci\^{e}ncias Naturais e Humanas, Universidade Federal do ABC, R. Santa Ad\'{e}lia 166, 09210-170 Santo Andr\'{e}, S\~{a}o Paulo, Brazil \\
$^{4}$Centro Brasileiro de Pesquisas F\'{\i}sicas, Rua Dr. Xavier Sigaud 150, 22290-180 Rio de Janeiro, Rio de Janeiro, Brazil}

\abstract{Nuclear Magnetic Resonance (NMR) was successfully employed to test several protocols and ideas in Quantum Information Science. In most of these implementations the existence of entanglement was ruled out. This fact introduced concerns and questions about the quantum nature of such bench tests. In this article we address some issues related to the non-classical aspects of NMR systems.  We discuss some experiments where the quantum aspects of this system are supported by quantum correlations of separable states. Such quantumness, beyond the entanglement-separability paradigm, is revealed via a departure between the quantum and the classical versions of information theory. In this scenario, the concept of quantum discord seems to play an important role. We also present an experimental implementation of an analogous of the single-photon Mach-Zehnder interferometer employing two nuclear spins to encode the interferometric paths. This experiment illustrate how non-classical correlations of separable states may be used to simulate quantum dynamics. The results obtained are completely equivalent to the optical scenario, where entanglement (between two field modes) may be present.}

\maketitle

\section{Introduction}

It is widely accepted that Nuclear Magnetic Resonance (NMR) is an experimental technique able to implement an ensemble quantum information processor using nuclear spins. From the creation of NMR pseudo-pure states, several quantum protocols for quantum information processing (QIP) were implemented in such physical system (Oliveira \emph{et al.}, 2007; Jones, 2011). NMR also has turned possible the experimental evaluation of abstract theoretical proposals in quantum information science as well as tests of principles in foundations of quantum mechanics (Auccaise \emph{et al.}, 2012), with a precision rarely obtained through other experimental techniques. Thus, NMR has became a powerful experimental approach to problems related to QIP, even though it has been criticized due to the absence of entanglement in most of the systems used up to now (Oliveira \emph{et al.}, 2007), besides the scalability issues (Jones, 2011).

The non-existence of entangled states became a problem when it was conjectured the necessity of such a resource to turn possible obtain an advantage in QIP (Braunstein \emph{et al.}, 1999). In this sense, as the states used to implement quantum protocols in NMR, named the pseudo-pure states, do not present entanglement except in singular situations, a questioning about the quantum nature of these implementations was raised (Laflamme \emph{et al.}, 2002). It was also shown, for a large class of protocols, that the existence of entangled states is a necessary (but not a sufficient) condition for quantum computers to present an exponential speed-up over the classical ones (Linden and Popescu, 2001), although too much entanglement could also be harmful (Gross \emph{et al.}, 2009; Morimae, 2010; Tilma \emph{et al.}, 2010). However, there are other important characteristics like the effectiveness of the implementation and the manipulation of quantum states, which also play some role in the quantum advantage game. Thus, considering that the principal aspect of NMR is the excellent control of unitary transformations promoted by radiofrequency (RF) pulses, such a spectroscopic technique allows one to obtain effective and singular methods in order to manipulate quantum states to implement QIP protocols. This is part of the reason of the well succeeded NMR-QIP experiments.

Notwithstanding, as suggested in Refs. (Laflamme \emph{et al.}, 2002; Vedral, 2010; Soares-Pinto \emph{et al.}, 2010; Auccaise \emph{et al.}, 2011a; Auccaise \emph{et al.}, 2011b), the presence of quantum correlations other than entanglement offers additional motivations for the successful NMR-QIP implementations, although, until now, not promoting the exponential speed-up of information processing. It is possible to measure such non-classical correlations in a bipartite system using a quantity called \emph{quantum discord} (Ollivier and Zurek, 2002; Henderson and Vedral, 2001) [for recent reviews about quantum aspects of correlations beyond the entanglement-separability paradigm see Refs. (C\'{e}leri \emph{et al.}, 2011; Modi \emph{et al.}, 2012)]. One interesting feature of such correlations is that some mixed-separable quantum states, or non-entangled states, can present non-null discord, indicating the existence of non-classical correlations beyond the entanglement-separability paradigm. These correlations may have a significant role in quantum information science due to the quantum advantage, when compared with its classical analogues, obtained by protocols based in non-entangled states (Vedral, 2010). These concepts allowed us to study experimentally and theoretically the existence of such correlations in NMR systems, as well as the effects of the decoherence (usually modelled as the amplitude- and phase-damping channels) over the system's non-classicality (Soares-Pinto \emph{et al.}, 2010; Auccaise \emph{et al.}, 2011a; Auccaise \emph{et al.}, 2011b; Soares-Pinto \emph{et al.}, 2011). Such a general kind of non-classical correlations is a source of quantumness (or a quantum resource) available in room-temperature NMR experiments.

In this article we present an NMR implementation of an analogous of the well-known single-photon Mach-Zehnder (MZ) interferometer employing two nuclear spins to encode the interferometric paths. MZ interferometer is an important and very useful tool for applications in quantum optics as well as for fundamental tests of quantum mechanics (Scully and Zubairy, 1997). Such a device enables us to determine phase shifts between two paths mediated by two half-silvered mirrors, called beam splitters (BS). At a single-particle level, the wave interference between the probability amplitude in the two arms of the interferometer plays a crucial role in the phase determination. It is worthwhile to note that it was believed that this interferometer, at a single quantum level, needed entanglement (between the field modes in both paths of the interferometer) to work properly, in other words, for a separable input state of the filed modes after the first beam splitter, the output should be also classical. In this case there is no interference between the two paths of a single quantum.  Experimental implementations of interferometry in NMR were already reported in the literature [see for example: (Peng \emph{et al.}, 2003; Peng \emph{et al.}, 2005; Auccaise \emph{et al.}, 2012)]. However, such experiments was realized using only one nucleus (one spin-$1/2$ particle) to encode the two interferometric paths. Here we present an analogous of a MZ interferometer by means of two nuclei that encodes the path information taken by a single quantum. This has some advantages, as discussed bellow, but our aim here is mainly to discuss the root of the non-classical aspects of NMR systems, as well as to give one more indicative of the potential of such a system as a quantum simulator.

The article is organized as follows. In Sec. \ref{Measures} we describe how one can quantify and measure quantum and classical correlations in highly mixed states in an NMR setup. Sec. \ref{Witness} is dedicated to the detection, but not quantification, of quantum correlations. The intention of these two sections is to emphasise the quantum nature of NMR systems, since one needs a quantum system to efficiently simulate another quantum system. The analogous of the single-photon MZ interferometer is implemented in Sec. \ref{Interferometer}. Finally, in Sec. \ref{Conclusions}, we present our final discussions.

\section{Measuring quantum and classical correlations in NMR}
\label{Measures}

For a typical liquid-state NMR system at room temperature, the gap between the Zeeman energy levels of the nuclei are much smaller than the average thermal energy, $\epsilon = \hbar \omega_{L}/2^{n}k_{B}T \sim 10^{-5}$ (with $\omega_{L}$ being the Larmor frequency), which implies that the density matrix can be written as the expansion (Abragam, 1978; Oliveira \emph{et al.}, 2007)
\begin{equation}
\rho\simeq\frac{1}{2^{n}}\mathbb{I}+\epsilon\Delta\rho,
\label{DM}
\end{equation}
where $\mathbb{I}$ is the $2^{n}\times2^{n}$ identity matrix, $n$ is the number of qubits (encoded in spin nuclei) and $\Delta\rho$ is the traceless deviation matrix. This is the well-known high-temperature approximation.

Any manipulation in the above state, like state preparation, quantum state tomography, qubit rotations and so on, is performed only over the deviation matrix $\Delta\rho$. This is due to the fact that such manipulations are, generally, sequences of RF pulses, which are basically unitary transformations $U$, that act on the density matrix in the following way
\begin{equation}
U\rho U^{\dag}=\frac{1}{2^{n}}\mathbb{I}+\epsilon U\Delta\rho U^{\dag}.
\label{ManipulDM}
\end{equation}
Adjusting each pulse length, phase and amplitude, it is possible to obtain a very fine control over the density matrix populations and coherences (diagonal and off-diagonal elements of the density matrix, respectively, usually defined in the computational basis). Together with proper temporal or spatial averaging procedures and evolution under spin interactions (Vandersypen and Chuang, 2004), the RF pulse can be specially designed to prepare any two-qubit computational base states, as well as its superpositions, starting from the thermal equilibrium state (Bonk \emph{et al.}, 2004; Bonk \emph{et al.}, 2005; Fahmy \emph{et al.}, 2008; Jones, 2011). Although being possible, in principle, to create any quantum state by means of these techniques, it was shown that the obtained state, in general, is a highly mixed one that does not possess entanglement since $\epsilon \sim 10^{-5}$, thus violating the boundary $\epsilon \geq 1/(1+2^{2\,n-1})$ for the existence of such correlations (Braunstein \emph{et al.}, 1999; Linden and Popescu, 2001).

Nevertheless, such separable state can present other kind of non-classical correlations that are conjectured to play a relevant role in QIP protocols (Datta \emph{et al.}, 2008; Lanyon \emph{et al.}, 2008; Vedral, 2010). The total correlation contained in a state $\rho_{AB}$ is quantified by the quantum mutual information (Nielsen \emph{et al.}, 2000; Benenti \emph{et al.}, 2007; Vedral, 2006)
\begin{equation}
\mathcal{I}(A:B)=S(A)+S(B)-S(AB),
\label{QMI}
\end{equation}
with $S(X) \equiv S(\rho_{X}) = -\mbox{Tr}\rho_{X}\ln \rho_{X}$ being the von Neumann entropy and $\rho_{A} = \mbox{Tr}_{B}\rho_{AB}$ being the reduced density operator for subsystem $A$, with an equivalent definition for partition $B$. This is a direct generalization of the classical mutual information, introduced by Shannon to quantify correlations in classical information theory: $I(A:B)=H(A)+H(B)-H(AB)$, with $H(A)=-\sum_{k}p^{A}_{k}\ln p^{A}_{k}$ being the Shannon entropy of the probability distribution $\{p^{A}_{k}\}$ for the random variable $A$ (Shannon, 1948). In the classical domain, we can rewrite this expression in the equivalent form $J(A:B)=H(A)-H(A|B)$, with $H(A|B)$ being the knowledge we can get from a random variable $A$ when we have measured $B$, i.e., the conditional entropy.  This is the point where the quantum and the classical domains breakup. While a classical measure does not disturb the system, a quantum one generally does. Therefore, the quantum extension of $J(A:B)$ is not straightforward, but a possible one  was considered in Ref. (Olliver and Zurek, 2002) as
\begin{equation}
\mathcal{J}(A:B) =S(\rho_{A})-S_{\left\{\Pi_{j}^{B}\right\}}\left(\rho_{A|B}\right),
\end{equation}
with $S_{\left\{\Pi_{j}^{B}\right\}}\left(\rho_{A|B}\right) = \sum_{j}q_{j}S(\rho_{A}^{j})$ being a quantum extension of the classical conditional entropy $\mathcal{H}(A|B)$. $\{\Pi_{j}^{B}\}$ is a complete set of projective measurements on partition $B$ and $\rho_{A}^{j} = \mbox{Tr}_{B}(\rho_{AB} \mathbf{1}_{A}\otimes\Pi_{j}^{B})/p_{i}$ ($p_{j} = \mbox{Tr}[\mathbf{1}_{A}\otimes\Pi_{j}^{B}\rho_{AB}]$) is the measured reduced density matrix of partition $A$. The difference
\begin{equation}
\mathcal{D}(A:B) = \mathcal{I}(A:B) - \max_{\left\{\Pi_{j}^{B}\right\}}\mathcal{J}(A:B)
\label{QD}
\end{equation}
was called quantum discord and is a measure of the quantumness of correlations (Olliver and Zurek, 2002, C\'{e}leri \emph{et al.}, 2011; Modi \emph{et al.}, 2012), since the expressions for $\mathcal{I}(A:B)$ and $\mathcal{J}(A:B)$ are classically equivalent. We note that the second term on the right hand side of Eq. (\ref{QD}) can be regarded as a measure of the classical correlations contained in the state $\rho_{AB}$ (Henderson and Vedral, 2001).

Let us back to the NMR system, where all correlations in the quantum state $\rho_{AB}$ comes from the deviation matrix $\Delta\rho_{AB}$. In this context, it is desirable to express the correlation quantifiers as functions of $\Delta\rho_{AB}$. In order to do this, we expand the von Neumann's entropy in powers of the parameter $\epsilon$ as 
\begin{equation}
S\left(\rho\right)  = 2\left(1-\frac{\epsilon^{2}}{\ln2}\mbox{Tr}\Delta\rho^{2}\right)  + \cdots , 
\label{Entro}
\end{equation}
where we have used $\mbox{Tr}\Delta\rho=0$. As the reduced-density operators reads $\rho_{A(B)}=\mbox{Tr}_{B(A)}\rho=\mathbf{1}_{A(B)}/ 2+\epsilon\Delta \rho_{A(B)}$, with $\Delta\rho_{A(B)}=\mbox{Tr}_{B(A)}\Delta\rho_{A(B)}$ being the reduced deviation matrix, the marginal entropies became
\begin{equation}
S\left(\rho_{A(B)}\right)  =1-\frac{\epsilon^{2}}{\ln2}\mbox{Tr}\Delta\rho_{A(B)}^{2}+\cdots . 
\label{EntroRed}
\end{equation}
Substituting Eqs. (\ref{Entro}) and (\ref{EntroRed}) into Eq. (\ref{QMI}) and keeping terms up to second order we obtain
\begin{equation}
\mathcal{I}\left(  \rho\right)  \simeq\frac{\epsilon^{2}}{\ln2}\left(2\mbox{Tr}\left(\Delta\rho\right)^{2}-\mbox{Tr}\left(\Delta\rho_{A}\right)^{2}-\mbox{Tr}\left(\Delta\rho_{B}\right)^{2}\right), 
\label{QMI_DM}
\end{equation}
which is the desired expression for total correlations.

To quantify the classical correlations we must obtain the measured density operator, which is given by (in a symmetric form) (C\'{e}leri \emph{et al.}, 2011; Modi \emph{et al.}, 2012)
\begin{eqnarray}
\eta &=& \sum_{i,j}\left(  \Pi_{i}^{A}\otimes\Pi_{j}^{B}\right)  \rho\left(\Pi_{i}^{A}\otimes\Pi_{j}^{B}\right) \nonumber\\
&\equiv& \frac{\mathbf{1}}{4}+\epsilon\Delta\eta , 
\label{DM_Proj}
\end{eqnarray}
where we defined the measured deviation matrix as $\Delta\eta\equiv\sum_{i,j}\left(  \Pi_{i}^{A}\otimes\Pi_{j}^{B}\right)  \Delta\rho\left(\Pi_{i}^{A}\otimes\Pi_{j}^{B}\right)$. Note that we applied the projection operators on both partitions $A$ and $B$. This is a symmetric version of the classical correlations defined in Eq. (\ref{QD}) and from here on we will adopt such definition to quantify correlations [see (C\'{e}leri \emph{et al.}, 2011; Modi \emph{et al.}, 2012) for details and discussions about non-classical correlation quantifiers].

Following the reasoning that led us to Eq. (\ref{QMI_DM}), we obtain the following expression for the mutual information of the measured state
\begin{equation}
\mathcal{J}\left(\eta\right) \simeq \frac{\epsilon^{2}}{\ln2}\left\{2\mbox{Tr}\left(\Delta\eta\right)^{2}-\mbox{Tr}\left(  \Delta\eta_{A}\right)^{2}  - \mbox{Tr}\left(\Delta\eta_{B}\right)^{2}\right\}, 
\label{QMMI}
\end{equation}
and thus for the classical correlation
\begin{equation}
\mathcal{C}\left(\rho\right) \simeq \max_{\left\{\Pi_{i}^{A}\otimes\Pi_{j}^{B}\right\}}\mathcal{J}_{c}\left(\eta\right), 
\label{CC}
\end{equation}
where $\Delta\eta_{A(B)}=\mbox{Tr}_{B(A)}\Delta\eta$. The quantum correlation in the composed two-qubit system can be directly computed from Eqs. (\ref{QMI_DM}) and (\ref{CC}) as (Soares-Pinto \emph{et al.}, 2010)
\begin{equation}
\mathcal{Q}(\rho)=\mathcal{I}\left(\rho\right)  -\mathcal{C}\left(\rho\right). 
\label{CQvf}
\end{equation}

\begin{figure}[h]
\begin{center}
\includegraphics[scale=0.33]{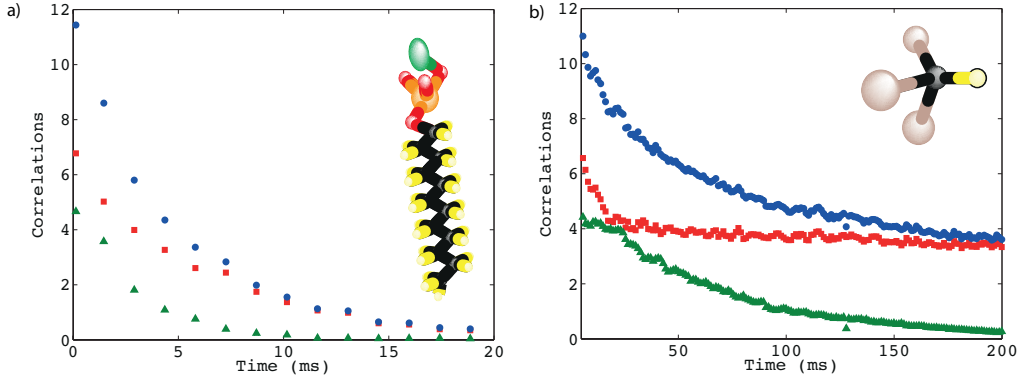}
\caption{(Online version in colour.) Relaxation dynamics of correlations. The blue circles are the mutual information while the red squares and green triangles represents classical and quantum correlations, respectively. In panel (a) we consider two logical qubits encoded in SDS molecule under the action of the same phase-damping environment while panel (b) shows the dynamics of the two physical qubits of the chloroform. The insets show a schematic diagram of the molecular structure of both systems. In both cases we have prepared the same initial state. For details see Refs. (Soares-Pinto \emph{et al.}, 2010; Auccaise \emph{et al.}, 2011a).}
\end{center}
\end{figure}

Equations (\ref{CC}) and (\ref{CQvf}) are general and can be used to quantify quantum and classical correlations in every system whose density matrix can be cast in the form given by Eq. (\ref{DM}). Specifically for NMR systems, we have employed such expansions to study the relaxation dynamics of correlations in two different system: ($i$) a sodium dodecyl sulfate (SDS) liquid-crystal sample containing $^{23}$Na quadrupolar nucleus with spin $I=3/2$ (Soares-Pinto \emph{et al.}, 2010) and ($ii$) a liquid sample of  $^{13}$C enriched chloroform, where we have two spin-$1/2$ systems  ($^{1}$H and $^{13}$C nuclei) (Auccaise \emph{et al.}, 2011a). We can use system ($i$) to encode two \emph{logical} qubits, while system ($ii$) encodes two \emph{physical} qubits. It is worthwhile to note that, during the decoherence dynamics, the amplitude-damping channel (which describes energy loss) acts individually on each qubit in both systems, while the phase-damping channel (responsible for coherence loss) is a global channel in case ($i$) and it acts independently over each qubit in case ($ii$) (Soares-Pinto \emph{et al.}, 2011; Souza \emph{et al.}, 2010). This difference allowed us to detect, in system ($ii$), a peculiar behaviour of the decoherence dynamics of quantum discord, i.e., a sudden change in its decay rate (Auccaise \emph{et al.}, 2011a) (theoretically predicted in (Maziero \emph{et al.}, 2009)). 

These results reveal two interesting aspects of NMR systems. First, there are quantum correlations in these systems, that can be quantified, for instance, by the quantum discord. Second, since the relaxation mechanisms present distinct features in both systems, they provide a very interesting platform to investigate the differences introduced in the dynamics by a global and a local environment. An example of this difference is illustrated in Fig. 1 where the relaxation dynamics of the correlations between two qubits is shown. While, in the case of a global reservoir (Fig. 1a), we have the expected exponential decay of all kinds of correlations (Soares-Pinto \emph{et al.}, 2010), the sudden-change behaviour clearly takes place in the case of independent reservoirs (Auccaise \emph{et al.}, 2011a), as can be seen in Fig. 1b (see the caption of the figure for details). Moreover, it is quite remarkable that the sudden-change still takes place even in the presence of the amplitude-damping channel, indicating that such phenomenon could be a strong characteristic of the correlations.

\section{Witnessing the quantumness of correlations in NMR}
\label{Witness}

In the last section we discussed the quantification of quantum and classical correlations in a composite systems. This is a very expensive information since a full quantum state tomography, beyond the numerical optimization process to compute Eq. (\ref{CC}), must be performed. However, there are many situations in which we do not need to know how much correlations a certain state possess, instead it is enough to know only its nature. In order words, we just have to distinguish between classical and quantum correlations. In the same spirit of what happens to entanglement, some non-classical correlation criteria and observable witnesses were proposed in Refs. (Rahimi and Saitoh, 2010; Girolami and Adesso, 2011; Brodutch and Modi, 2011; Zhang \emph{et al.}, 2011; Yu \emph{et al.}, 2011; Maziero and Serra, 2010) and experimentally verified in Refs. (Auccaise \emph{et al.}, 2011b; Passante \emph{et al.}, 2011;  Aguilar \emph{et al.}, 2012). A classicality (or non-classicality) witness is regarded as an observable (or a set of observables) that can be directly measured in an experimental setup. Depending on its expected value, we know if the state has quantum or only classical correlations. The measurement of such a witness in an NMR system revealed directly the non-classical nature of its highly mixed state (Auccaise \emph{et al.}, 2011b; Passante \emph{et al.}, 2011).

In what follows, we describe a non-classicality witness and its experimental measurement performed in an NMR setup at room temperature (Auccaise \emph{et al.}, 2011b). In Ref. (Maziero and Serra, 2010) it was shown that the mean value of the operator
\begin{equation}
W_{\rho}=\sum_{i=1}^{3}\sum_{j=i+1}^{4}\left|\langle O_{i}\rangle_{\rho}\langle O_{j}\rangle_{\rho}\right|=0, \label{witness}
\end{equation}
is a sufficient condition for classicality of correlations for a wide class of two-qubit systems, expressed by
\begin{equation}
\rho=\frac{\mathbb{I}^{AB}}{4}+\frac{1}{4}\sum_{i=1}^{3}(\mathcal{A}_{i}\sigma_{i}^{A}\otimes\mathbb{I}^{B}+\mathcal{B}_{i}\mathbb{I}^{A}\otimes \sigma_{i}^{B}+\mathcal{C}_{i}\sigma_{i}^{A}\otimes \sigma_{i}^{B}). 
\label{2qubits}
\end{equation}
Here $O_{i}=\sigma_{i}^{A}\otimes \sigma_{i}^{B}$ for $i=1,2,3$ and $O_{4}=\sum_{i=1}^{3}(z_{i}\sigma_{i}^{A}\otimes\mathbb{I}^{B}+w_{i}\mathbb{I}^{A}\otimes \sigma_{i}^{B})$. $\sigma_{i}^{A(B)}$ is the $i$th Pauli operator acting on subsystem $A(B)$. $\mathcal{A}_{i},\mathcal{B}_{i}, \mathcal{C}_{i}, z_{i}, w_{i} \in \Re$, with $z_{i}$ and $w_{i}$ randomly chosen and constrained such that $\sum_{i}z_{i}^{2}=\sum_{i}w_{i}^{2}=1$.

In the case of Bell-diagonal class of states
\begin{equation}
\rho_{bd}=\frac{\mathbb{I}^{AB}}{4}+\frac{1}{4}\sum_{i=1}^{3}\mathcal{C}_{i}\sigma_{i}^{A}\otimes \sigma_{i}^{B}, \label{BellDiag}
\end{equation}
$W_{\rho_{bd}} = 0$ is also a necessary condition for the absence of quantumness in the correlations of the composite system (for this case  $\langle O_{4}\rangle_{\rho_{bd}}=0$ (Maziero and Serra, 2010)).

One interesting fact about such witness is that it is possible to rewrite the observables in Eq. (\ref{witness}) in terms of one component of the magnetization in one subsystem as $\langle O_{i}\rangle_{\rho}=\langle\sigma_{1}^{A}\otimes\mathbb{I}^{B}\rangle_{\xi_{i}}$, with $\xi_{i}=U_{A\rightarrow B}[R_{n_{i}}(\theta_{i})\rho R_{n_{i}}^{\dagger}(\theta_{i})]U^{\dagger}_{A\rightarrow B}$, where $R_{n_{i}}(\theta_{i})=R_{n_{i}}^{A}(\theta_{i})\otimes R_{n_{i}}^{B}(\theta_{i})$, being $R_{n_{i}}^{A(B)}(\theta_{i})$ a local rotation by an angle $\theta_{i}$ around direction $n_{i}$ on subsystem $A(B)$, where $\theta_{1}=0$, $\theta_{2}=\theta_{3}=\pi/2$, $n_{2}=y$. And $n_{3}=z$, $U_{A\rightarrow B}$ is the CNOT gate with subsystem $A$ being the control qubit. This fact leads to a straightforward implementation of this witness in the NMR scenario, since the one-qubit magnetizations are the natural observables for these systems. In fact, the witness in Eq.  (\ref{witness}) was experimentally implemented in a room temperature NMR two-qubit system (Auccaise \emph{et al.}, 2011b), directly revealing the non-classical aspects of highly mixed states. 

\section{Mach-Zehnder interferometer}
\label{Interferometer}

An interesting way to test non-classicality in NMR systems can be provided by an analogous of the well-known single-photon Mach-Zehnder interferometer. In order to perform such an interferometer and test the role of correlations, we employ two nuclei to encode the two interferometric path information. As already mentioned, this approach differs significantly from the previous interferometric measures implemented in the NMR scenario (Auccaise \emph{et al.}, 2012; Peng \emph{et al.}, 2005; Peng \emph{et al.}, 2003), in which the two-path information are encoded in just one nuclear spin. Moreover, due to the fact that distinct nuclei have, generally, distinct relaxation times, it turns possible the study of the environment-induced phase shift between both paths. This environment-induced phase shift may found applications, for example, in thermometry (Stace, 2010) and in the quantum illumination protocol (Lloyd, 2008).

Now, let us briefly review the single-photon MZ interferometer in the optical scenario, schematically shown in Fig. 2. In this figure, $BS$ is a $50:50$ beam splitter (one half-silvered mirror), $M$ is a mirror and $D_{A}$ and $D_{B}$ are one-photon sensitive detectors. The phase difference $\phi$ between paths $A$ and $B$ can be due to an environment-induced phase or just a controlled one.  In what follows, the indexes $A$ and $B$ must be understood as two spatial field modes (referring to distinct paths taken by the photon).

\begin{figure}[h]
\begin{center}
\includegraphics[scale=0.4,angle=90]{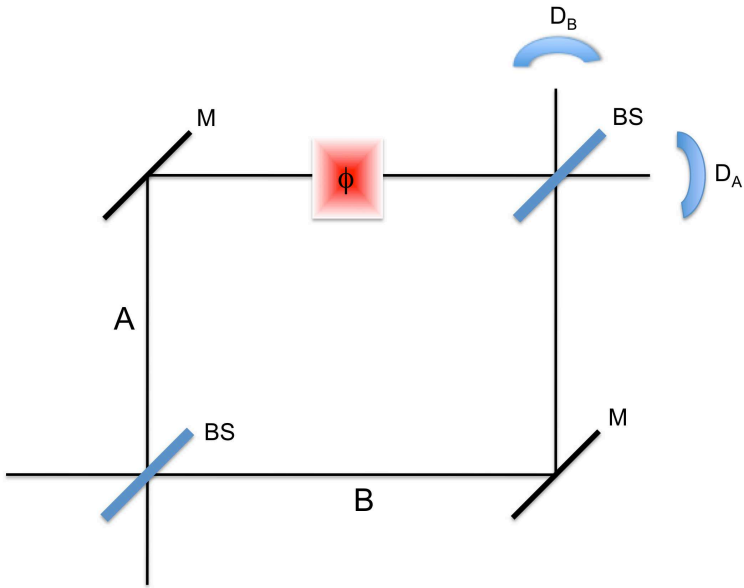}
\caption{(Online version in colour.) Schematic diagram of the Mach-Zehnder interferometer.}
\end{center}
\end{figure}

Let us suppose that the input (pure) state (before the first BS in Fig. 2) is given by
\begin{equation}
\vert \Psi_{i} \rangle = \vert 0 \rangle_{A}\otimes\vert 1 \rangle_{B}.
\label{init}
\end{equation} 

The state after the first beam splitter is
\begin{equation}
\vert \Psi \rangle = \frac{1}{\sqrt{2}}\left(\vert 0 \rangle_{A}\otimes\vert 1 \rangle_{B} + i\vert 1 \rangle_{A}\otimes\vert 0 \rangle_{B}\right).
\end{equation}
Note that this can be seen as an entangled state of the two field modes (that encode path information). The coherent superposition of both paths after the first beam splitter is essential for the interferometer to produce an interference pattern (Scully and Zubairy, 1997). In our experiment, the information about both paths will be encoded in two distinct nuclei and, i.e., when talking about correlations between these nuclei, we mean the coherent superposition paths $A$ and $B$.

The final state, after the second BS (see Fig. 2), will be given by
\begin{equation}
\vert \Psi_{f} \rangle = \cos\left(\frac{\phi}{2}\right)\vert 0 \rangle_{A}\otimes\vert 1 \rangle_{B} - \sin\left(\frac{\phi}{2}\right)\vert 1 \rangle_{A}\otimes\vert 0 \rangle_{B}.
\label{final}
\end{equation}

From here on we will omit the tensor product symbol to simplify notation. Equation (\ref{final}) shows the interference pattern between the two paths and reflects the wave-like behaviour of the one-photon state. The probability to detect the photon in the detector $D_{B}$ (the probability of detect the initial state at the end of the interferometer) is given by $cos^{2}\left(\phi/2\right)$. If we introduce a non-destructive detector in any of the paths before the second beam splitter, this interference pattern disappears and we observe the particle character of the quantum system. This is one the most known representations of Bohr's complementarity principle. It is interesting to note that we could modify the present scheme to quantitatively study this principle by means of an inequality relating the which-way information (particle)  and the fringe visibility of the interferometer (wave), as suggested in (Englert, 1996).

We implement such interferometer employing a two-qubit system comprised by nuclear spins of $^1$H and $^{13}$C atoms in a carbon-13 enriched chloroform molecule (CHCl$_{3}$). The sample was prepared by mixing $50$ mg of $99$\% $^{13}$C-labelled CHCl$_{3}$ in $0.7$ ml of $99.8$\%  CDCl$_{3}$ in a 5 mm NMR tube. Both samples were provided by the Cambridge Isotope Laboratories - Inc. The experiments were performed at 25$^{\circ}$ C using a Varian $500$ MHz Premium Shielded ($^{1}$H frequency), at the Brazilian Centre for Physics Research. A Varian $5$ mm double resonance probe-head equipped with a magnetic field gradient coil was used.  

In our NMR version of the MZ interferometer, we have exactly the same situation of the optical standard version. The two paths are encoded in two different nuclei, with just one quantum of excitation. The nuclear spin in the ground state represents the vacuum field state, while the nuclear  spin in the excited state represents the field excitation (the one photon state). We choose the $^{13}$C nucleus to encode the path in which we apply the controlled phase (path $A$ in Fig. 2), while the $^1$H nucleus encodes the reference path (path $B$ in Fig. 2). This choice was motivated by the fact that the Carbon nucleus has the smaller transversal relaxation time [see (Auccaise \emph{et al.}, 2011a; Auccaise \emph{et al.}, 2011b) for details about the relaxation times for our system]. It is not important for the present experiment, but can be explored by experiments studying the effects of decoherence in just one path, like the thermometry mentioned above (Stace, 2010). Here, the phase difference will be provided by a controlled $z$-rotation applied to the Carbon nucleus.

The pulse sequence used to implement the analogous of the MZ interferometer is shown in Fig. 3. The first block of this figure is used to prepare the deviation matrix in a state analogous to $\vert\Psi\rangle = \vert 0\rangle\vert 1\rangle$, which means that the Hydrogen nucleus is in the ground state while Carbon is in its excited state. Some words about the meaning of this analogy are in order. In Sec. \ref{Measures} we have mentioned that the spin state is represented by the highly mixed density operator shown in Eq. (\ref{DM}). This happens because the magnetic energy and the thermal energy ratio is $\epsilon \sim 10^{-5}$, so the density operator was decomposed in two contributions: the white noise term and the deviation matrix term. Recalling that the control over the system is given by RF pulses, which are represented by unitary operations (besides gradient pulses), we only act in the deviation matrix term (second term in Eq. (\ref{DM})). We then encode all the information of the state $\vert\Psi\rangle$ in this matrix.  At the end we will have a quantum dynamics equivalent to that of a pure state.  

\begin{figure}[h]
\begin{center}
\includegraphics[scale=0.5]{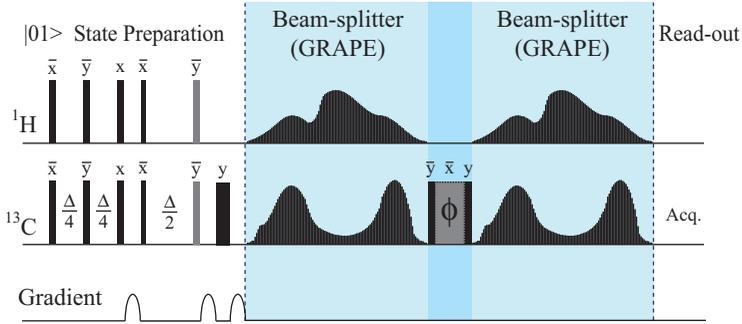}
\caption{(Online version in colour.) Pulse sequence for the implementation of the analogous of a single-photon Mach-Zehnder interferometer in a two nuclei NMR system. The thicker filled bars represent $\pi/2$ pulses, the thinner bars indicate $\pi/4$ pulses, and the grey bars indicate $\pi/6$ pulses with the phases as shown (negative pulse phases are described by a bar over the symbol). $\Delta = 1/J$ and the time periods in the state preparation represent free evolutions under the $J$ coupling (Oliveira \emph{et al.}, 2007).}
\end{center}
\end{figure}

After the state preparation, there are three consecutive blocks that are used to implement the MZ interferometer. The first and the third ones play the role of the beam splitters. We have constructed a sequence of 128 pulses with modulated frequencies and amplitudes by means of the GRAPE optimization method (Machnes \emph{et al.}, 2008) to implement the following action on both qubits
\begin{equation}
\mbox{BS} = \frac{1}{\sqrt{2}}\left[
\begin{array}{cccc}
1 & 0 & 0 & 0 \\
0 & 1 &  i & 0 \\
0 & i &  1 & 0 \\
0 & 0 &  0 & 1
\end{array}
\right], 
\end{equation}
which represents the beam splitter operation. The time duration of the entire sequence in order to implement such operation is $4.5$ms. Finally, the three pulses between the beam splitters implement a rotation on the Carbon nucleus about the $z$ axis by an angle $\phi$. After the pulse sequence depicted in Fig. 3 we performed a full quantum state tomography. We repeated the protocol in Fig. 3 for different phase shifts.

The experimental results are presented in Fig. 4, where we plot the probability of detecting the initial state $P_{i} = \vert \langle \psi_{i}\vert \psi_{f}\rangle\vert^{2}$ as a function of the phase shift $\phi$. The theoretical prediction from Eqs. (\ref{init}) and (\ref{final}) are in a good agreement with the experimental results showing that the interference expected for the interferometer can be obtained in the context of the highly mixed states NMR system. It is remarkable that the tiny coherence present in our experiment (of order $\epsilon$) enables the interferometer to work properly. The visibility of the interferometer (which quantifies the contrast of the interference in any system which has wave-like character) is near to unity, without entanglement between the two paths (the two nuclear spins in this case). We may argue that the reason for this is the quantum correlation of separable states (quantified by the quantum discord) present in the composite nuclear spin system. 

The small deviation between the observed and the expected probability in Fig. 4 is mainly due to the fact that we are continuously applying a RF pulse on the Carbon spin to obtain the controlled phase shift between both paths. This pulse sequence presents some fluctuations introducing noise into the system. Moreover, the pulse is not exactly in resonance with the qubit transition frequency, which contributes to the phase mismatch between theory and experiment. We observe that there are two decoherence processes involved here. The first one is that caused by the usual amplitude- and phase-damping channels, which are always present, but that have a negligible effect for the current experiment due to the fact that it occurs in a time much shorter than the characteristic relaxation times of the system. All the observed decoherence is due to the imperfections of the RF pulses employed to generate the phase difference $\phi$ between both paths.

\begin{figure}[h]
\begin{center}
\includegraphics[scale=0.5]{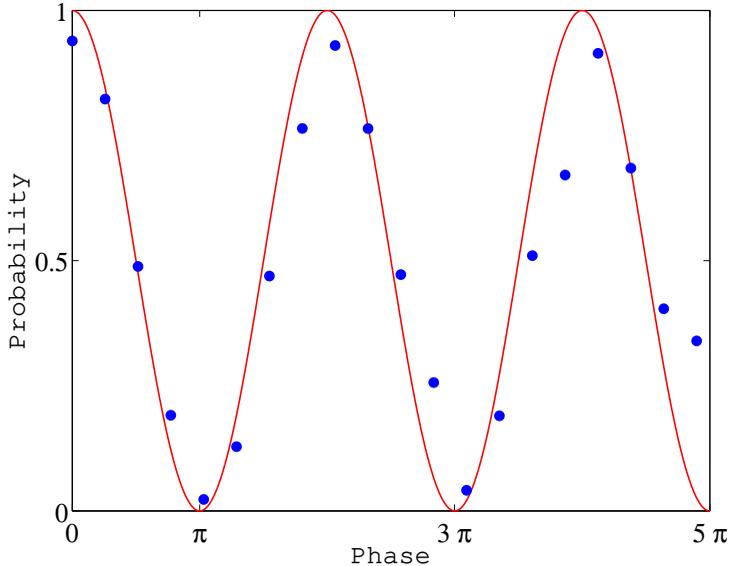}
\caption{(Online version in colour.) Probability for the detection of the initial state in the experimental implementation of the Mach-Zehnder interferometer in a two nuclei NMR system. The (blue) dots are the experimental results and the (red) solid line is the theoretical prediction of Eqs. (\ref{init}) and (\ref{final}).}
\end{center}
\end{figure}

To give some support to our statement that quantum correlations of separable states are the responsible for the successful simulation of the single-photon interferometer, in Fig. 5 we plot the visibility and the quantum discord between both qubits (paths) in a slight modified experiment, where we vary the amount of quantum discord after the first BS operation in the pulse sequence of Fig. 3. To obtain this result we performed two experiments. In the first one, we let the system to freely evolve (under the natural decoherence) during a certain period of time $\tau$ just after applying the first BS operation and, then, we perform a full quantum state tomography. From the obtained deviation matrix we compute the quantum discord through Eq. (\ref{CQvf}). By varying the time $\tau$ of the decoherent evolution we obtain different amounts of quantum discord, the results are shown by the red squares in Fig. 5. In the second experiment we again, after the first beam splitter, let the system to freely evolve under the action of the environment. However, instead of performing the state tomography, we proceed with the sequence shown in Fig. 3 and compute the fringe visibility of the interferometer
\begin{equation}
\mathcal{V} = \frac{\max\langle\Psi_{f}\vert\hat{D}_{A}\vert\Psi_{f}\rangle - \min\langle\Psi_{f}\vert\hat{D}_{B}\vert\Psi_{f}\rangle}{\max\langle\Psi_{f}\vert\hat{D}_{A}\vert\Psi_{f}\rangle + \min\langle\Psi_{f}\vert\hat{D}_{B}\vert\Psi_{f}\rangle},
\end{equation}
with $\hat{D}_{A}$ and $\hat{D}_{B}$ being the detectors operators. As we can see from Fig. 5, the decay of the discord is clearly accompanied by the decay in the visibility, showing that, without discord, we cannot obtain a visible interference pattern from the simulated MZ interferometer.

\begin{figure}[h]
\begin{center}
\includegraphics[scale=0.5]{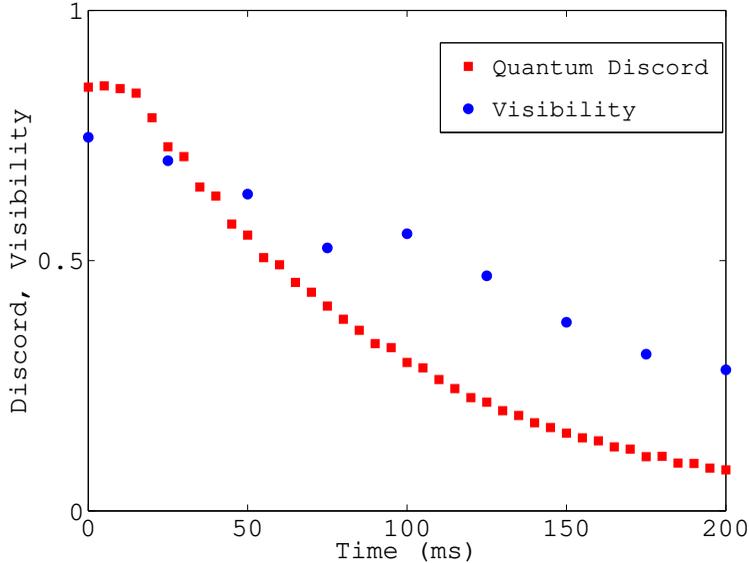}
\caption{(Online version in colour.) Quantum discord and visibility of the MZ interferometer for different decoherence periods $\tau$. Note that, although both quantities are represented in the same vertical axis, they are completely different quantities. While the visibility is a dimensionless quantity, the quantum discord is measures in unities of $\varepsilon^{2}/\ln 2$ bit. The purpose of the graph is only to show that, the fade of the interference pattern is accompanied by the vanishing of the quantum discord between both qubits (paths of the interferometer).}
\end{center}
\end{figure} 

\section{Final Discussions}
\label{Conclusions}

The QIP either for computing or communication, probably will lead to a technological revolution in this century. Having appeared in early 1980s, Quantum Computation and Quantum Information evolved faster as an area of basic research in physics and as a promise technological alternative. In fundamental physics, the concepts of quantum information has stimulated a huge number of new ideas and  results, which have deepened our knowledge about the nature of the quantum behaviour. Particularly exciting are the relations of quantum correlations and natural resources for computation and communication. The discovery of quantum algorithms, new forms of communication, and cryptography led to a remarkable experimental development in order to test these applications.

Up to now, almost all quantum algorithms and communication protocols have been tested on systems with a few qubits. However, for large-scale applications of quantum devices it is necessary to study the generation, handling and the storage of quantum states in such systems. We may say that the experimental technique that have stood out in this context was NMR, because it allows a quite precise manipulation of quantum states. This precise control of spins states is due to the radio-frequency technology developed over decades. It has been used in an inventive way for NMR methods and contributed to the impressive and fast success of the technique in QIP (Oliveira \emph{et al.}, 2007; Jones, 2011). NMR also allows, through the manipulation of nuclear spin qubits, the simulation of more complex systems (\'{A}lvarez and Suter, 2010; Du \emph{et al.}, 2010; Lu \emph{et al.}, 2011; Souza \emph{et al.}, 2008; Souza \emph{et al.}, 2011).

Although all the successful use of NMR in QIP, some years ago it was shown the impossibility of existence of entangled states in such experimental technique at room temperature. By that time, it was supposed that this kind of quantum correlation was intrinsically linked with the speed-up of the quantum algorithms. Thus, one question could be raised: What is the quantum aspect of NMR states that allows quantum dynamics and its application in QIP? This question is partially answered by the existence of quantum correlations in separable states, as revealed, for example, by the quantum discord. As we have argued in this article, such kind of correlations seems to occur naturally in room temperature NMR systems, as far as it can be easily quantified or witnessed by the methods discussed here, and it is responsible for the quantumness of the system. The analogous single-photon MZ interferometer presented in the last section provides a clear illustration of the non-classicality present in NMR highly mixed states. The coherence present in the deviation matrix allows for a path interference analogous to that one obtained in the optical context where an entangled state (between two field modes) is present. In other words, we may argue that the separable state (non-entangled) quantum correlations present in the two qubit NMR state are the resource that enables the interferometer to work properly with a visibility near to unity. 

In summary, the phase coherence between two-nuclear spins in the NMR highly mixed states can encompass non-classical correlations (as measured by the quantum discord). Such quantum correlation are available for performing quantum dynamics (path interference) allowing the simulation of several systems, as for instance, the single-photon MZ interferometer. The discussions presented here may opens the way for a new and very exciting tests of the quantum aspects of nature and of protocols in information science that exploit this kind of non-classicality, which is present even in the NMR highly mixed states scenario.   

\ack{The authors acknowledge financial support from UFABC, CAPES, and FAPESP. This work was performed as part of the Brazilian National Institute for Science and Technology on Quantum Information (INCT-IQ). LCC also acknowledge the pleasant time spent in the NMR Laboratory of the Brazilian Center for Physics Research, during which part of this work was done and some others were initiated in the company of all members of the laboratory, usually also accompanied by a good beer.}

\end{document}